# Nature of electrons from oxygen vacancies and polar catastrophe at LaAlO$_3$/SrTiO$_3$ interfaces

Xiaorong Zhou and Zhiqi Liu*

School of Materials Science and Engineering, Beihang University, Beijing 100191, China

E-mail: zhiqi@buaa.edu.cn

**Abstract**

The relative significance of quantum conductivity correction and magnetic nature of electrons in understanding the intriguing low-temperature resistivity minimum and negative magnetoresistance of the two-dimensional electron gas at LaAlO$_3$/SrTiO$_3$ interfaces has been a long outstanding issue since its discovery. Here we report a comparative magnetotransport study on amorphous and oxygen-annealed crystalline LaAlO$_3$/SrTiO$_3$ heterostructures at a relatively high-temperature range, where the orbital scattering is largely suppressed by thermal fluctuations. Despite of a predominantly negative out-of-plane magnetoresistance effect for both, the magnetotransport is isotropic for amorphous LaAlO$_3$/SrTiO$_3$ while strongly anisotropic and well falls into a two-dimensional quantum correction frame for annealed crystalline LaAlO$_3$/SrTiO$_3$. These results clearly indicate that a large portion of electrons from oxygen vacancies are localized at low temperatures, serving as magnetic centers, while the electrons from the polar field are only weakly localized due to constructive interference between time-reversed electron paths in the clean limit and no signature of magnetic nature is visible.

## 1. Introduction

The discovery of the two-dimensional electron gas (2DEG) at the interface of two large bandgap insulators LaAlO$_3$ (LAO) and SrTiO$_3$ (STO) [1] has spurred a gold rush for unveiling exotic functionalities at strongly correlated oxide interfaces [2] for more than one decade. As a model two-dimensional (2D) oxide interface system, a plethora of physical phenomena [3-5] such as the coexistence of ferromagnetism and superconductivity [6-9] have been demonstrated for the LAO/STO interface, which has thus received significant attention from the entire condensed matter physics society.

Among many attractive features of the 2DEG, a low-temperature resistivity minimum typically occurs in LAO/STO heterostructures fabricated at relatively high oxygen pressures, which was first observed in 2007 by Brinkman *et al.* and explained by magnetic Kondo scattering resulting from localized Ti 3*d* moments [10]. Many studies also used the Kondo effect to explain the low-temperature electron scattering for LAO/STO [11-16]. However, the magnetotransport studies performed by Caviglia *et al.* [17] and other studies [18-23] suggested that instead of magnetic scattering, the low-temperature electron scattering at the LAO/STO interface was dominated by quantum conductivity correction due to the constructive interference between time-reversed electron paths, *i.e.*, weak localization.

Up to now, low-temperature electron scattering mechanism for electrons at the LAO/STO interface has been controversial [10-26]. One important aspect on this divergence could be that the origin of electrons at the LAO/STO interfaces can be different from group to group depending on the fabrication conditions. It was found that in crystalline LAO/STO (c-LAO/STO) heterostructures that are not oxygen annealed upon deposition, a large portion of free electrons originates from oxygen vacancy doping as a result of high chemical affinity of Al atoms to oxygen on the surfaces of STO substrates [27]. After oxygen annealing, most of electrons from oxygen vacancies could be removed and hence electrons from the polar catastrophe mechanism are left. In contrast, in amorphous LAO/STO (a-LAO/STO) heterostructures which are fabricated by room-temperature deposition of LAO onto STO substrates, the 2DEG solely results from oxygen vacancies and fully vanishes after oxygen annealing [27,28].

Electrons originating from oxygen vacancies and the electronic reconstruction could be substantially distinct in terms of the degree of localization especially at low temperatures. For example, oxygen-vacancy-induced electrons tend to be localized to the defect level while the temperature is lowered. That is because the oxygen vacancy defect level is separated by 4-25 meV [27,29] from the bottom of the conduction band of STO, consequently leading to the carrier freeze-out effect which has been extensively observed in oxygen-vacancy-dominated STO-based electron systems [6,10,27-32]. On the contrary, electrons originating from the polar catastrophe mechanism are directly transferred from the LAO valence band to the STO conduction band as a result of the electric potential build-up within the polar LAO layer while growing on non-polar STO. Consequently, the carrier density of this mechanism is almost independent of temperature [27,33,34]. From this perspective, it is rather necessary to separate the two types of electrons and investigate their low-temperature scattering properties separately.

## 2. Results and Discussions

To achieve this goal, 4-nm-thick a-LAO/STO heterostructures were fabricated from a single crystal LAO target on untreated (100)-oriented STO substrates by pulsed laser deposition system (Shenyang Baijujie Scientific Instrument Co., Ltd) with base pressure of $1.5\times10^{-8}$ Torr and $10^{-3}$ Torr oxygen pressure at room temperature. The 10-uc-thick c-LAO/STO heterostructures were fabricated on $TiO_2$-terminated (100)-oriented STO substrates at 750°C and $10^{-3}$ Torr oxygen pressure, and the growth of these samples were monitored *in situ* by reflection high-energy electron diffraction (RHEED). Subsequently these samples were oxygen-annealed at 600°C under 1 atm of oxygen for 2 h to remove most of the oxygen vacancies. The electrical measurements were performed at a Quantum Design physical properties measurement system with its resistivity option. As shown in Fig. 1a, the room-temperature sheet resistance of an annealed c-LAO/STO heterostructure is one order of magnitude higher than that of an a-LAO/STO heterostructure. Nevertheless, below 50 K, they exhibit close sheet resistance. The zoom-in plots for temperatures below 30 K (Figs. 1b & c) show a similar resistivity minimum at ~10 K for both heterostructures.

As shown in Fig. 2a, the sheet carrier density of the a-LAO/STO heterostructure is ~$8\times10^{13}$ cm$^{-2}$ at 300 K, which is rather comparable with the previously reported values for oxygen-vacancy-dominated 2DEG (2DEG-*V*) [27,28,31,32]. In contrast, the room-temperature carrier density for the annealed c-LAO/STO is much lower, ~$1.6\times10^{13}$ cm$^{-2}$, consistent with that of the polar-catastrophe-induced 2DEG (2DEG-*P*) in annealed c-LAO/STO samples studied before [27,33,34]. An obvious carrier freeze-out effect is seen below ~150 K for the 2DEG-*V* in the a-LAO/STO sample while the carrier density of the 2DEG-*P* in the annealed c-LAO/STO sample possesses a rather weaker temperature dependence, which is in good agreement with previous studies [27,38,31-34]. The temperature dependence of carrier mobility $\mu$ (Fig. 2b) is similar for the two types of 2DEG. Power law dependences are obvious above 30 K, which indicates the predominant role of phonon scattering [35]. The change of the power law dependence at ~105 K is a feature of the STO structural phase transition.

To minimize the effect of Lorentz orbital scattering, which is typically proportion to the square of mobility, and meanwhile avoid the suppression of the magnetoresistance (MR) effect by thermal fluctuations, we chose a moderate temperature 30 K to first examine magnetotransport properties of the two types of heterostructures. As plotted in Fig. 3a, the MR of the 2DEG-*V* in the a-LAO/STO heterostructure is all negative when the magnetic field is applied out-of-plane, parallel to the measuring current, and in-plane normal to the current. More importantly, the MR is almost the same for all the three different measurement geometries. This is largely different from the weak localization induced negative MR for a 2D electron system, which is absent when the magnetic field is applied within the 2D plane [36]. Instead, it is consistent with the magnetic Kondo scattering scenario. In this case, localized electrons from oxygen vacancies lead to localized Ti 3*d* moments, which serve as magnetic scattering centers. When a magnetic field is applied along any direction, the localized moments are aligned and thus the magnetic scattering of electrons is reduced, leading to negative MR. In addition, the isotropic MR indicates the absence of magnetic anisotropy and thus long-range magnetic interaction among localized moments is not expected. The Kondo effect usually leads to a resistivity minimum at low temperatures as the magnetic scattering dominates over the impurity scattering and becomes stronger at lower temperatures, which is in accordance with our experimental observation in Fig. 1a.

In sharp contrast, the MR of the 2DEG-*P* in the annealed a-LAO/STO heterostructure is strongly anisotropic (Fig. 3b). The negative MR is only significant when the magnetic field is applied out-of-plane while the MR effect almost vanishes when the magnetic field is applied in-plane. This agrees well with the weak localization mechanism for a 2D transport system. That is because the quantum conductivity correction for time-reversed electron paths can only occur when the magnetic field is perpendicular to the 2D electron orbital plane in the weak localization picture. Moreover, the negative MR in the weak localization is typically small, on the order of one percent [24], which is consistent with our experiment result.

To further confirm the weak localization mechanism of the negative MR for the 2DEG-*P*, we measured the out-of-plane MR effect at different temperatures. As shown in Fig. 4a, the magnetoconductance $\delta G = G(B) - G(0)$ for the 2DEG-*P* at various temperatures data well fits the Hikami-Larkin-Nagaoka formula

$$\delta G = \frac{\alpha e^2}{2\pi^2 \hbar}\left[\psi\left(\frac{1}{2} + \frac{\hbar}{4eBl_\phi^2}\right) - \ln\left(\frac{\hbar}{4eBl_\phi^2}\right)\right]$$ [37-39]

where $\psi(x)$ stands for the digamma function, $l_\phi$ is the phase coherence length and $\alpha$ represents a parameter depending on the number of conduction channels, which decides whether the quantum conduction correction belongs to weak localization ($\alpha > 0$) or weak antilocalization ($\alpha < 0$). Via fitting, we obtain positive values for α at all the temperatures (Fig. 4b), which hence proves that the weak localization is the main mechanism for the negative MR. This naturally explains the resistivity minimum observed in the temperature-dependent sheet resistance for the annealed c-LAO/STO sample (Fig. 1a). It is worthy noticing that the negative MR in our samples persists up to 9 T from 50 K to 30 K, which indicated that the weak localization effect persists up to the magnetic field of 9 T. This result agrees with some typical low-dimensional conduction systems [17,40,41]. However, in some other low-dimensional conduction systems [42,43], the MR is negative for moderate magnetic field, followed by a positive MR at higher magnetic field, which indicated that some other effect like regular orbital scattering may become more important at higher magnetic field. The dominating factors for the difference in low-dimensional conduction systems need to be further investigated.

The mean free path $l_{MF}$ for the 2DEG-$P$ sample can be estimated via $l = \frac{h}{\sqrt{2\pi n}e^2 R_{Sheet}}$, where $h$ is the Planck constant, $n$ is the sheet carrier density and $R_{Sheet}$ is the sheet resistance. As plotted in Fig. 4b, both the mean free path $l_{MF}$ and the phase coherence length $l_\phi$ decrease with increasing temperature due to enhanced thermal fluctuations. However, $l_{MF}$ is obviously larger than $l_\phi$ of the weak localization for all the studied temperatures. This illustrates that the 2DEG-$P$ system in our case belongs to the clean limit so that the weak localization mechanism is not suppressed by scattering mechanisms.

The distinct magnetotransport properties for the two types of electrons originating from oxygen vacancies (2DEG-$V$) and the polar catastrophe (2DEG-$P$) at LAO/STO interfaces clearly indicate that the magnetic Kondo scattering exists for the 2DEG-$V$ while the weak localization is dominant for 2DEG-$P$. This can be well understood by different degrees of localization for the two types of electrons. Specifically, most of electrons from oxygen vacancies in STO become localized at low temperatures (Fig. 2a), giving rising to localized Ti 3$d$ moments, which is consistent with theoretical study that oxygen vacancies at LAO/STO interfaces enhance magnetic moments [44]. While the electrons in the polar catastrophe mechanism transferred from the LAO valence band to the STO conduction band are not very sensitive to the temperature. At low temperatures, thermal fluctuations are suppressed and thus the quantum interference effect between time-reversed electron paths becomes stronger for a 2D transport system, resulting in only weak localization.

## 3. Conclusions

In summary, via magnetotransport properties studies of the two types of LAO/STO heterostructures that exhibit a resistivity minimum at low temperatures, we have clarified that the magnetic Kondo scattering is the dominant mechanism for the resistivity minimum for the oxygen-vacancy-induced 2DEG while the weak localization accounts for the resistivity minimum for the polar-catastrophe-induced 2DEG. This work resolves the controversy that has been outstanding in this field for many years. In addition, the clear pictures revealed by this study could be useful for understanding other exotic physical phenomena in this interface system, *e.g.*, the coexistence of magnetism and superconductivity [6-9], and the electron transport in other 2D systems.

## Acknowledgements

Z.L. acknowledges financial support from the National Natural Science Foundation of China (NSFC; grant numbers 11704018, 51822101, 51861135104 & 51771009).

**Figure 1.** (a) Temperature-dependent sheet resistance for a 4-nm-thick amorphous LAO/STO (a-LAO/STO) heterostructure and a 10-uc-thick oxygen annealed crystalline LAO/STO (c-LAO/STO) heterostructure. (b) Low-temperature sheet resistance for the a-LAO/STO sample. (c) Low-temperature sheet resistance for the annealed c-LAO/STO sample.

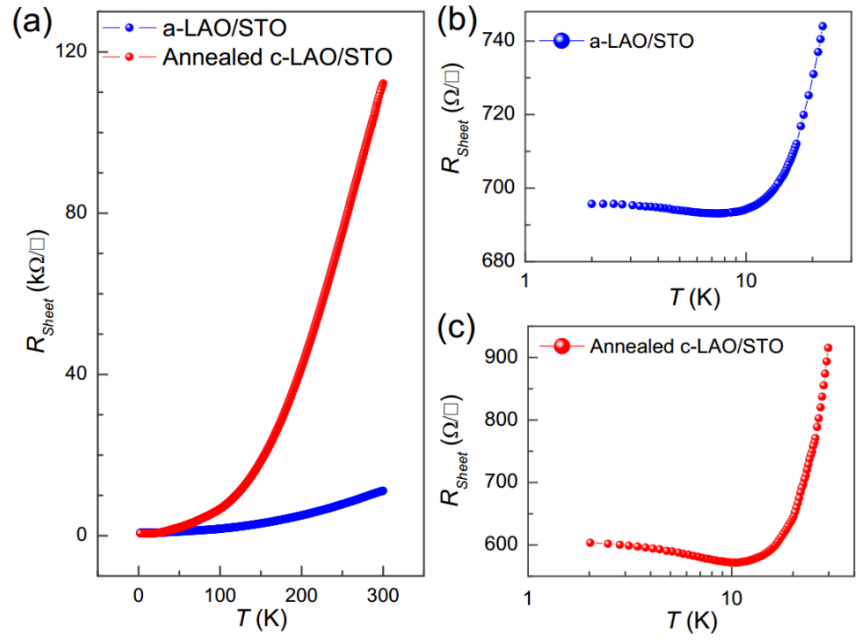

**Figure 2.** (a) Temperature-dependent sheet carrier density for the two types of LAO/STO heterostructures. (b) Temperature-dependent mobility for the two types of LAO/STO heterostructures and power law fittings.

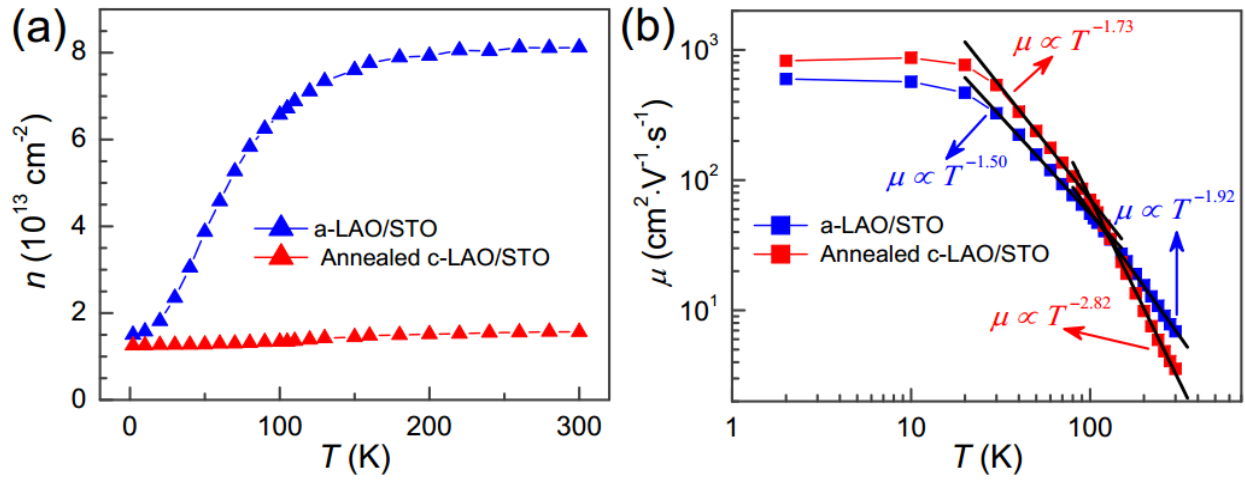

**Figure 3.** (a) Magnetoresistance (MR) of the a-LAO/STO sample at 30 K under different measurement geometries. (b) MR of the annealed c-LAO/STO sample at 30 K under similar measurement geometries.

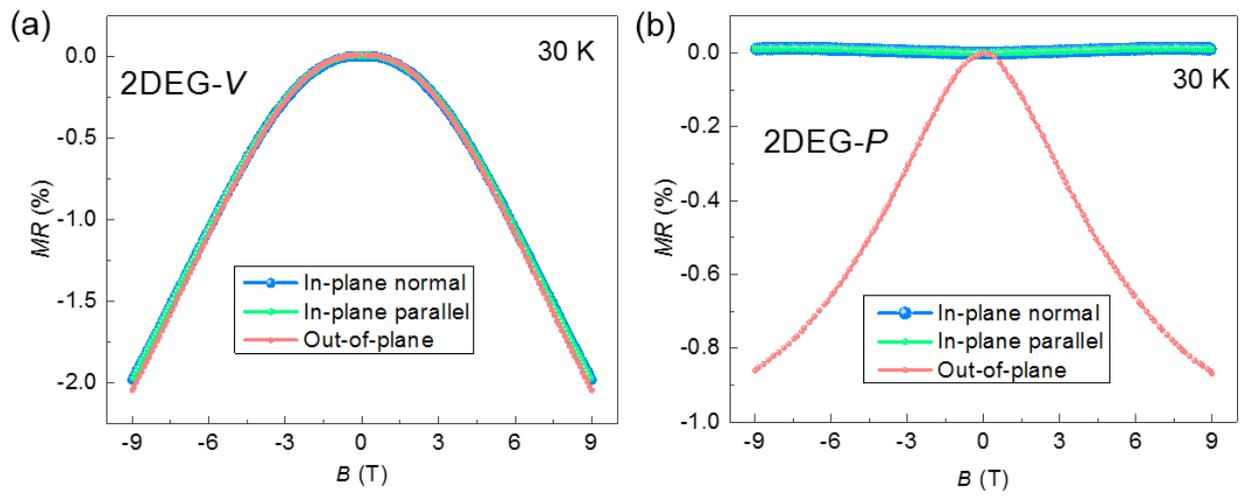

**Figure 4.** (a) Magnetoconductance of the annealed c-LAO/STO heterostructure at different temperatures. The circles represent measured data points and the solid lines are fitted curves. (b) Fitting parameter, phase coherence length and mean free path as a function of temperature.

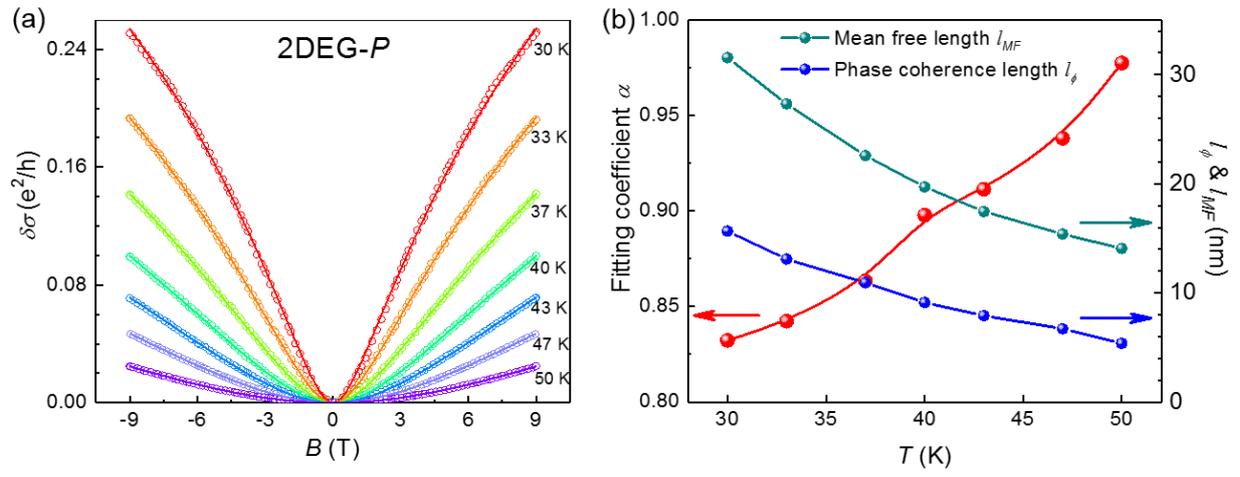